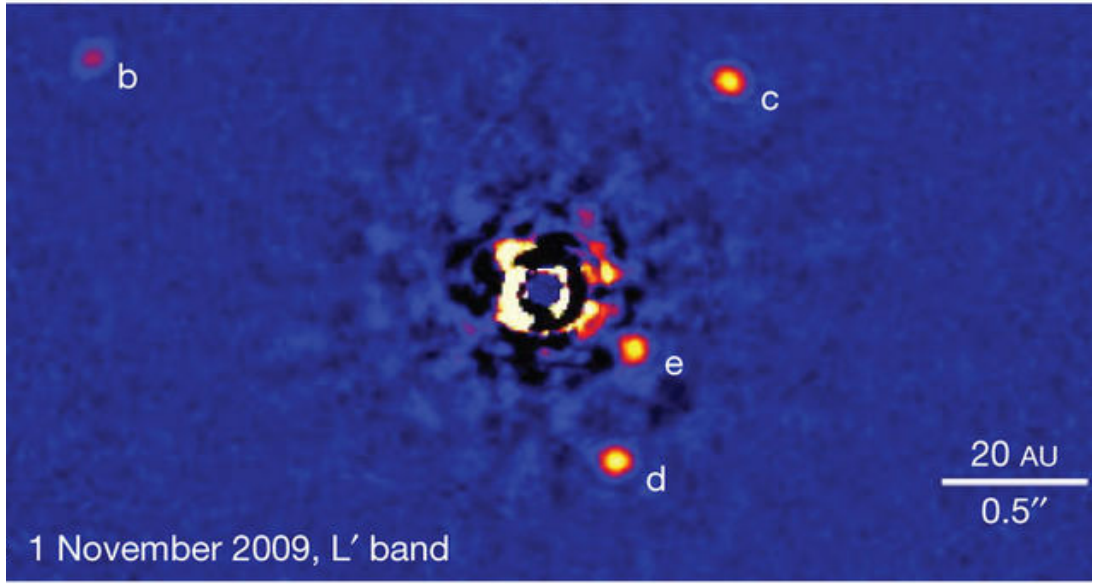

*The first directly-imaged planetary system, HR 8799, discovered with Keck and Gemini.*

# Using Ground-Based Telescopes to Mature Key Technologies and Advance Science for Future NASA Exoplanet Direct Imaging Missions

Lead Author: Thayne Currie          NASA-Ames Research Center/Subaru Telescope
                                    email: currie@naoj.org, phone: (857) 998-9771

Co-Authors:   Ruslan Belikov         NASA-Ames Research Center
              Olivier Guyon          Subaru Telescope/University of Arizona
              N. Jeremy Kasdin       Princeton University
              Christian Marois       NRC-Herzberg
              Mark S. Marley         NASA-Ames Research Center
              Kerri Cahoy            Massachusetts Institute of Technology
              Michael McElwain       NASA-Goddard Spaceflight Center
              Eduardo Bendek         NASA-Ames Research Center
              Marc J. Kuchner        NASA-Goddard Spaceflight Center
              Michael R. Meyer       University of Michigan

Co-Signees:   Yasuhiro Hasegawa (JPL), Wladimir Lyra (CSUN/JPL), Bertrand Mennesson (JPL), Chris Packham (UT-San Antonio)

**Summary:** Ground-based telescopes have been playing a leading role in exoplanet direct imaging science and technological development for the past two decades and will continue to have an indispensable role for the next decade and beyond. Extreme adaptive optics (AO) systems will advance focal-plane wavefront control and coronagraphy, augmenting the performance of and mitigating risk for *WFIRST-CGI,* while validating performance requirements and motivating improvements to atmosphere models needed to unambiguously characterize solar system-analogues with *HabEx/LUVOIR*. Specialized instruments for Extremely Large Telescopes may deliver the first thermal infrared images of rocky planets around Sun-like stars, providing *HabEx/LUVOIR* with numerous exo-Earth candidates and key ancillary information that can help clarify whether the planets are habitable.

**1. Background: The Ground's Historical Role in Advancing Exoplanet Direct Imaging**

Since the first nearby extrasolar planets were discovered from indirect detection methods [1], NASA has shown a strong interest in someday directly imaging and spectrally characterizing solar system-like planets, including an Earth twin around a Sun-like star (e.g. the *Terrestrial Planet Finder* mission; [2]).

Nearly two decades later, though, ground-based telescopes coupled with adaptive optics (AO) systems have accounted for nearly all of the 15 to 20 exoplanets and candidates imaged thus far (e.g., [3], see [4] for summary). Follow-up photometry and spectra of these first discoveries provided insights into young super-Jovian planets' fundamental atmospheric properties: clouds, carbon chemistry, and surface gravity [5,6,7]. Some recent exoplanet direct imaging discoveries – enabled by *extreme* AO, which achieves deeper contrasts than facility systems – have revealed cooler, lower-mass planets with far different spectra [8].

Ground-based telescopes has also demonstrated and matured key technologies and methods forming the backbone of future NASA direct imaging missions. Extreme AO systems have demonstrated the capability to create a sub-arcsecond dark hole via high-order deformable mirrors (DM) driven by advanced wavefront sensors [9,10]. The ground has demonstrated advanced coronagraph designs like the vector vortex [11]. Sophisticated post-processing and spectral extraction methods were developed from analyzing ground-based data [12,13] and demonstrated to improve planet imaging capabilities from space [14]. Ground-based data clarified how speckle noise statistics relate to spectrophotometric uncertainties and affect planet detection limits [15,16,17].

The 2010 Decadal survey included technology development for a future direct imaging missions in its top-ranked medium scale priority. Undoubtedly, laboratory-based investments in wavefront control and coronagraphy have significantly advanced this goal for *WFIRST-CGI* (e.g. [18]). Detailed simulations have also better optimized the performance of future NASA flagship missions utilizing larger segmented mirrors, such as *HabEx* or *LUVOIR* (e.g. [19,20]).

As described below, in addition to the lab, the ground will also continue to play an indispensable role for maturing key technologies and advancing science for future space missions.

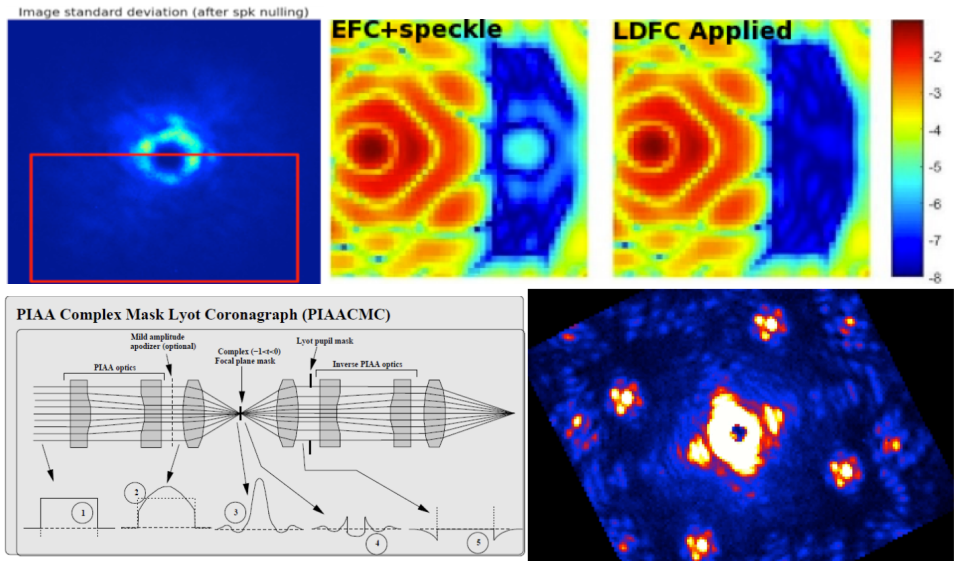

*Figure 1 - Key technologies matured on the ground and applicable to future NASA missions. (Top-Left) On-sky demonstration of speckle nulling from [22]. (Top-Right) Simulation demonstrating Linear Dark Field Control (LDFC) [27]. (Bottom-Left) schematic of the PIAACMC coronagraph [28] and (bottom-right) SCExAO/CHARIS data for HR 8799 using the shaped-pupil coronagraph: LDFC, PIAACMC and other advances (e.g. Multi-Star Wavefront Control) will also be matured on SCExAO.*

## 2. Maturing Technologies Through Ground-Based Extreme AO

*WFIRST-CGI* will include components like high-order focal-plane wavefront control driven by methods such as electric field conjugation (EFC), low noise detectors for science and wavefront sensing, and advanced coronagraphy designed to further suppress starlight within several λ/D. *HabEx/LUVOIR* will require matured versions of these components [21]. In the next decade, ground-based extreme AO systems will demonstrate and clarify how these technologies improve scientific output when integrated into a larger system architecture that is used to detect and extract physical parameters from exoplanets (Fig. 1).

Keck/NIRC2 and Subaru/SCExAO provided the empirical demonstrations of the speckle nulling focal-plane wavefront control method [22, 23]. P1640 explored EFC's sensitivity to DM actuator responsiveness and pupil stop and focal plane mask alignments [24,25]. Ultra-low noise MKIDS detectors provide a sensitive science camera and fast focal-plane wavefront sensing [26] that may help extreme AO systems like GPI, SCExAO, and SPHERE deliver deeper raw (post-processed) contrasts of $10^{-6}$ ($10^{-7}$–$10^{-8}$) at small angles in the near-infrared (near-IR).

The ground is also now honing newer wavefront control methods potentially applicable to future NASA missions. For example, Subaru/SCExAO will help mature *Linear Dark Field Control* (LDFC), which utilizes the linear response of the region outside a dark hole dug by EFC/speckle nulling (the ``bright field'' or BF) to correct wavefront perturbations that affect both the BF and the DF [27]. While methods like EFC use DM probes to update the estimate of the electric field which perturb the science exposure, LDFC freezes the DH state initially achieved with EFC, potentially allowing greater observing efficiency and deeper contrasts.

Ground-based extreme AO systems will continue to mature new coronagraph designs relevant for future NASA missions. The PIAACMC design will be used on upgraded SCExAO and MagAO-X systems delivering near-IR Strehl ratios in excess of 90% and utilize both high-order focal-plane wavefront sensing and coronagraphic low-order wavefront sensing. On-sky tests of PIAACMC will help assess how the design's sensitivity to low-order aberrations, the trade-off between its suppression at small angles vs. bandwidth, etc. affect science capabilities.

Ground-based tests of new wavefront control and coronagraphy may reveal information hidden from laboratory predictions. For example, tests of the shaped-pupil coronagraph (SPC) with extreme AO assessed its sensitivity to low-order aberrations for Strehl ratios relevant for both extreme AO and future NASA missions [29]. While qualitatively the SPC performed as expected (low sensitivity to low-order modes), the on-sky data revealed that the relative advantage of the SPC over other designs may be larger than expected from laboratory tests alone.

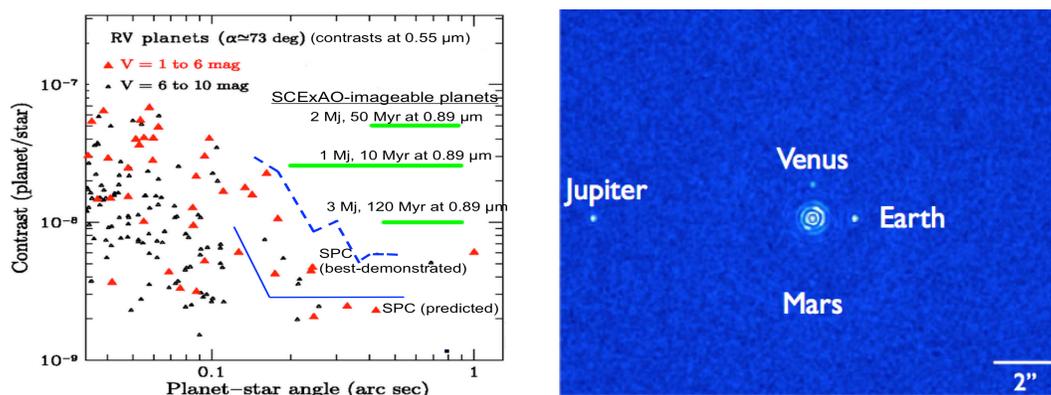

*Figure 2 – (Left) Even if WFIRST-CGI fails to image mature planets in reflected light, ground-based extreme AO systems enhanced by new focal-plane wavefront control methods like LDFC could provide a large sample of cool Jovian planets to spectrally characterize (see text for details). (Right) Simulated TMT/MICHI 10 μm image of the inner solar system at 4.3 ly.*

## 3. Advancing The Scientific Potential Of NASA Direct Imaging Missions With Ground-Based Extreme AO And ELTs

*WFIRST-CGI* will combine focal-plane wavefront control/coronagraphy with planet spectral extraction in space for the first time. *HabEx/LUVOIR* will leverage on this demonstration. In both cases, suitable direct imaging targets, improved planet atmosphere models, a deep understanding of uncertainties in spectral extraction, and constraining ancillary data are key for mission success. Current ground-based extreme AO systems and the upcoming generation of Extremely Large Telescopes (ELTs) help to offer these critical elements (Fig. 2).

For example, even a partial implementation of new wavefront control methods/coronagraphy that improve current extreme AO systems' contrasts at 0.1" – 0.5" by a factor of ~10 instead of ~100 (~$10^{-6}$–$10^{-7}$) could allow the detection of 1--3 $M_J$, 10 –120 Myr old planets, which are just 20—30 fainter in the red optical and recoverable with *WFIRST-CGI* (Marley et al. 2018, in prep.). They probe a phase space poorly explored in planetary atmospheres: 375–500 K objects near the T to Y dwarf transition but at low gravity. Like the first directly-imaged exoplanets, these newly-characterized exoplanets will challenge existing models and lead to a richer understanding of key atmospheric properties like clouds and chemistry but at temperatures and masses more characteristic of solar system planets (see white paper by Marley and Lewis). Thus, even if WFIRST-CGI fails to deliver the contrasts needed to image mature reflected-light Jovian planets ($\sim 10^{-9}$), ground-based extreme AO could provide the mission younger Jupiters detectable in thermal emission to demonstrate and advance precise atmosphere parameter estimation.

The new phase space in contrast that may be probed using current extreme AO systems and those on ELTs (e.g. the planned *Planetary Systems Imager* on TMT) may also help clarify the performance requirements needed to unambiguously characterize solar system-like planets with future NASA missions. At contrasts characteristic of the current ground-based extreme AO systems ($\sim 10^{-4}$–$10^{-6}$) relatively achromatic phase errors dominate the speckle noise budget. At contrasts closer to *WFIRST-CGI*'s performance ($10^{-7}$–$10^{-9}$) chromatic errors should begin to dominate. This change in regime affects post-processing and the spectral covariance of integral field spectrograph data, which impacts the signal-to-noise ratio needed for a given atmospheric characterization goal [17]. Future ground-based systems reaching $10^{-7}$–$10^{-8}$ contrast at several $\lambda/D$ probe a regime where speckle statistics are more uncertain [see also 16] and may provide helpful input about spectral retrieval and required signal-to-noise for atmosphere characterization goals in the extremely faint speckle regime in which future NASA missions will operate.

While space may be needed to directly detect an Earth twin around a Sun-like star in *reflected light*, ELTs may be able to image exo-Earths in *thermal* emission at 10 μm. At this wavelength, the habitable zone for Sun-like stars lies beyond 3 $\lambda/D$ for TMT and E-ELT out to 5–6 pc. The E-ELT/METIS and TMT/MICHI instruments couple low-emissivity adaptive optics delivering an exceptionally high-Strehl correction with advanced coronagraphy, yielding contrasts at r < 1" separations of $\sim 10^{-7}$ in several hours of integration time. Direct imaging searches for rocky planets receiving the same insolation as the Earth are possible for METIS and MICHI around ε Eridani, τ Ceti, and over a half-dozen other nearby ~solar-mass stars. The discovery space for rocky planets receiving Mars-like insolation (50% larger $\lambda/D$) or Jovian planets is even larger.

Thermal infrared imaging of rocky planets with ELTs also substantially increases the science gain of NASA direct imaging missions. Disentangling the effect of a planet's radius from its albedo based on reflected-light spectra alone is extremely challenging. However, thermal infrared data helps to constrain the equilibrium temperature of rocky planets and in turn the planet radius. Multi-epoch 10 μm imaging data will help identify optimal times for reflected-light space observations. Using the ground to pre-select the best exo-Earth candidates for follow-up in general could increase the yield for exo-Earth detection by *HabEx/LUVOIR* and may be especially advantageous for missions employing a starshade for spectroscopic follow-up.

### 4. Recommendations

The 2010 Decadal Survey noted the importance of ground-based research for constraining exoplanet demographics (i.e. with indirect methods) and typical levels of exozodiacal light through the nulling interferometers (e.g. with the LBTI/HOSTS program). While the discussion of direct imaging technology development focused entirely on laboratory-based advances, ground-based facilities have served and will continue to serve as key incubators for direct imaging technologies and sources of critical preparatory/complementary data for space missions.

We first recommend that NASA take an explicit interest in ground-based developments in focal-plane wavefront control, coronagraphy, advanced post-processing/spectral retrieval methods, and the practical gains achieved by integrating these components.   One option is to support more formal partnering between the laboratory and observatory.   Doing so allows for not just a cross-pollination of ideas benefiting both NASA and ground-based observatories in the near term but helps the next generation of instrument scientists hone their skills so that they can advance NASA direct imaging missions in the long term.

Second, we recommend that ground-based direct imaging with extreme AO on 8-10m class telescopes and later on ELTs be explicitly considered as critical and necessary support for NASA's direct imaging-focused missions.   For example, precision radial-velocity measurements have been emphasized as key ground support for *WFIRST-CGI* as they provide jovian planets the mission can detect and characterize.   Extreme AO upgraded with new wavefront control and coronagraphy over the next decade will do the same **without** requiring that WFIRST achieves ~ $10^{-9}$ contrast.   Hosts for these planets will also include early type stars (e.g. Vega, Fomalhaut, Altair) that provide *WFIRST-CGI* with a much more photon-rich environment for wavefront control and whose planets are inaccessible to precision radial-velocity detection.   Dedicated *WFIRST-CGI* precursor surveys on 8–10 m class telescopes operating with extreme AO could identify a large sample of WFIRST-accessible planets in thermal emission.   If ~ $10^{-8}$ contrast is reached, some planets in reflected light are reachable.   Planets suitable for *WFIRST-CGI* follow-up may also be imageable with ELTs at 3–5 μm (see white paper by M. Meyer et al.).   Surveys at 10 μm with ELTs will likewise precede and provide critical support for identifying habitable rocky planets around Sun-like stars with *HabEx/LUVOIR*.